\documentstyle[11pt,fleqn]{article}
\begin{document}
\baselineskip=5.5mm

\newcommand{\HI}{H\,{\sc i}\,}
\newcommand{\HII}{H\,{\sc ii}\,}
\newcommand{\um}{$\mu$ m }

{\Large \bf
\noindent Enhanced Star Formation in Seyfert 2 Galaxies}

\noindent
Q.-S. Gu, J.-H. Huang, and L. Ji

{\it 
\noindent
Department of Astronomy, Nanjing University, Nanjing 210093, China \\
\noindent
United Lab for Optical Astronomy, The Chinese Academy of Sciences, China
}

\vspace*{5.mm}
\noindent
{\bf Abstract}

\noindent
 In this paper, we report our preliminary results on enhanced star formation
 activity in Seyfert 2 galaxies. By re-analysing the Tully-Fisher relation for
 Whittle(1992)'s sample and for a Seyfert 2s' sample selecting 
 from Veron-Cetty \& Veron (1996), we find that 
 (1) almost all Seyfert 2 galaxies with circumnuclear star formation have a 
 ratio of far infrared (FIR) to blue luminosities ($\rm L_{\rm FIR}/L_{\rm B}$)
 to be larger than 1/3; (2) for Seyfert 2 galaxies with $\rm L_{\rm FIR}/L_{\rm B}
 < 1/3$, the Tully-Fisher relation is similar to that of the normal spiral galaxies;
 while for those with $\rm L_{\rm FIR}/L_{\rm B} \geq 1/3 $, they
 are significantly different from the normal ones, which 
 confirms Whittle's suggestion of enhanced star formation activities
 in the circumnuclear regions of these Seyfert 2 galaxies.
 
\vspace*{3.mm}

\vspace*{5.mm}
\section{Introduction}

 There are many debates on the connection between Seyfert activity and 
 starburst in AGN regime. On one hand, in the view of
 the starburst model for AGN, the Seyfert activity could be accounted for
 by a central starburst (Terlevich \& Melnick 1985 and Terlevich et al. 1992),
 on the other hand, Norman \& Scoville (1988) have suggested that a very compact starburst
 would provide mass to feed the nuclear black hole ( see also Perry \& Williams 1994 ).
 Recently, the UV-optical observations of four Seyfert 2 galaxies 
 (Heckman et al. 1997; Gonzalez Delgado et al. 1998) present
 direct evidence of the existence of hot, massive stars (O and B) in the
 nuclei of these Seyferts. In MKN 477, the contribution by 
 a nuclear starburst ( in the Wolf-Rayet phase) to the total intrinsic UV 
 luminosity can be about 25 \% (Heckman et al. 1997).
 Wilson (1988) suggested that a number of Seyfert galaxies had low ionization
 rotating gas disks and the high ionizion Narrow Line Regions (NLRs), such as 
 NGC 1068, NGC 7469, MKN 509 etc.,
 which showed clear evidence for circumnuclear star formation at both radio and IR wavelengths.
 Now, the growing observed data provided strong evidence indicating the coexistence of Seyfert 
 activity and massive starburst (e.g. Telesco et al. 1984; Rodriguez-Espinosa et al., 1987;
 Heckman  et al. 1989; Neff et al. 1994; Genzel et al. 1995; Oliva et al. 1995;
 Cid Fernander \& Terlevich 1995, and Gonzalez Delgado et al. 1997a).

 Colina et al. (1997b) have found that the observed UV fluxes in Seyfert 2 galaxies with 
 circumnuclear star-forming ring, are dominated {\it not} by the radiation from AGN, 
 but by emission from clusters of young hot stars located in circumnuclear star-forming 
 regions.  Now, more and more observational evidences suggest that nuclear starburst plays a 
 key role in Seyfert 2 nuclear activity ( e.g. Rodriguez-Espinosa et al., 1986; 
 Heckman et al. 1995; Heckman et al. 1997;
 Maiolino et al. 1995; and Gu et al. 1997), 
 
 In examining the basic acceleration mechanisms on the ionized gas in Seyfert's NLRs,
 Whittle (1992a,b,c) has suggested that the departures of Tully-Fisher and 
 Faber-Jackson relations in Seyfert galaxies from those of normal spirals
 are due to enhanced star formation in Seyferts. 
 The offset of Faber-Jackson relation is clearly
 explained by Nelson \& Whittle (1996), who have found that Seyfert
 bulges are kinematically normal, and the offset is due to a lower
 mean mass-to-light ratio of Seyfert galaxies, either having
 higher star formation rate
 or avoiding systems with older stellar populations. In this paper,
 we report our study on the offset of Tully-Fisher relation of Seyfert
 galaxies to normal spirals and  confirm the Whittle's suggestion of
 enhanced star formation in Seyferts, by use of the ratio of far-infrared (FIR)
 to blue luminosities ($\rm L_{\rm FIR}/L_{\rm B}$) as an indicator of recent
 star formation rate (SFR).
  
 The organization of this paper is as follows. 
 In section 2, we discuss $\rm L_{\rm FIR}/L_{\rm B}$ as an 
 indicator of star formation rate. 
 Our analyses on the Tully-Fisher relation of 
 Seyfert galaxies for Whittle's sample and for a sample of all Seyfert 2 galaxies 
 in Veron-Cetty \& Veron (1996) are presented in section 3. 
 And a summary of our main results is given in section 4.
 
 \section{$\rm L_{\rm FIR}/L_{\rm B}$ : an Indicator of SFR}

 As is well known, $\rm H\alpha$ emission is a good indicator of massive star
 formation rate (SFR) in galaxies ( see recent review by Kennicutt 1998 ). 
 However, there are only limited number of 
 galaxies with detected $\rm H\alpha$ emission fluxes (
 Kennicutt \& Kent 1983; Romanishin 1990; Ryder \& Dopita 1994; 
 Young et al. 1996; Gonzalez Delgado et al. 1997a and Feinstein 1997) 
 Instead, Devereux \& Young (1991) have suggested that FIR emission is also 
 an indicator of SFR, by assuming that FIR emission arises from thermal
 re-radiation of dust heated by massive O, B stars.  In three nearest 
 galaxies (M31, M81, and M101), Devereux and his collaborators
 have found a good spatial correspondence between the
 deconvolved FIR images and $\rm H\alpha$ emission (Devereux et al. 1994;
 Devereux et al. 1995 and Devereux \& Scowen 1994). Other
 evidence includes the tight correlation between FIR and $\rm H\alpha$ 
 emission fluxes (Lonsdale Persson \& Helou 1987; Devereux \& Young 1990). 
 In fact, FIR emission is widely used as an intensive
 indicator of SFR in analysing the 
 infrared properties of galaxies (e.g. Keel 1993; Combes et al 1994; 
 Helou \& Bicay 1993 ; Kandalian 1997).
 In his recent review of star formation in galaxies, 
 Kennicutt (1998) has suggested that FIR emission can be a sensitive
 tracer of the young stellar population and SFR, and presented the diagnostic 
 methods used to measure SFRs including FIR luminosities. 

 The relation between FIR and radio fluxes is well known to be the best 
 correlation for extragalactic objects. Condon, Anderson \& Helou (1991)
 have found that Q, the ratio of infrared to radio fluxes, increases
 away from a nearly constant value as the galaxy changes from heating by
 young stars to heating by old stellar populations, 
 Helou \& Bicay (1993) claim that the constant Q is in 
 force for $\rm L_{\rm FIR}/L_{\rm B} \geq 1/3$.  Recently,
 in analysing the bar-enhanced star-forming activities in spiral galaxies. 
 Huang et al. (1996) have taken $\rm L_{\rm FIR}/L_{\rm B}$
 as an indicator of relative star formation rate 
 Interesting,
 they also find a critical value of $\rm L_{\rm FIR}/L_{\rm B} \sim 1/3$, 
 and the bar's enhanced effect on star formation activity is detectable 
 only for galaxies with $\rm L_{\rm FIR}/L_{\rm B} \geq 1/3$. 

 FIR emission in Seyfert galaxies can be accounted for by
 (1) thermal re-radiation of warm dust heated in star forming
 regions; (2) emission associated with an AGN, 
 either nonthermal flux coming directly from AGN or dust reradiation of
 nonthermal UV-optical continuum emission from the accretion disk 
 (Rowan-Robinson 1987). But more and more work performed in UV, optical,
 mid-infrared, far-infrared and radio wavelengths, supports that most of FIR 
 emission in Seyfert 2 galaxies originates from dust heated 
 by hot, massive (O and B) stars formed in the circumnuclear starburst region
 (Rodriguez-Espinosa et al. 1987; Dultzin-Hacyan et al. 1988; Vaceli 
 et al. 1993; Gu et al. 1997; and Rodriguez-Espinosa \& Perez Garcia 1998). 

 Table 1 lists all Seyfert 2 galaxies with detectable nuclear/circumnuclear 
 star formation that we collect from literatures. It is very interesting to
 notice that nearly all these galaxies have  $\rm L_{\rm FIR}/L_{\rm B} \geq 1/3$
 except NGC 5347. 
 According to Gonzalez Delgado et al. (1997b), the numbers of \HII regions
 in the inner and outer part of the disk are 63/46, 28/18, 98/19, 59/105 and 1/32
 for NGC 1068, NGC 1667, NGC 3982, NGC 5427 and NGC 5347, 
 respectively. In NGC 5347, there is only {\it one} \HII region in the inner 
 disk, which might account for the low value of $\rm L_{\rm FIR}/L_{\rm B}$,
 if the majority of dust responsible for FIR emission is located in the nuclear
 region. 

\vspace*{2.mm}
\noindent
Table 1. Seyfert 2 Galaxies with Circumnuclear Star Formation 

\noindent
\begin{tabular}{llll}
\hline
\hline
Name & $\rm L_{\rm FIR}/L_{\rm B}$ &  Type & Notes \\
\hline
NGC  262   &  0.465            & .SAS0*.   & GD97 \\
NGC 1068   &  1.050            & RSAT3..   & GD97,CG97b \\
NGC 1365   &  0.872            & .SBS3..   & FN98 \\
NGC 1667   &  0.690            & .SXR5..   & GD97,CG97b \\
NGC 3982   &  0.420            & .SXR3*.   & GD97,CG97b \\
NGC 4303   &  0.472            & .SXT4..   & CG97a \\
NGC 4945   &  0.639            & .SBS6*    & FN98 \\
NGC 5135   &  1.565            & .SBS2..   & GD98 \\
NGC 5347   &  0.259            & PSBT2..   & GD97 \\
NGC 5427   &  0.450            & .SAS5P.   & GD97,CG97b \\
NGC 5728   &  0.444            & .SXR1*.   & CA96 \\
NGC 5953   &  1.910            & .SA.1*P   & CG97b \\
NGC 6221   &  0.335            & .SBS5..   & FN98 \\
NGC 6810   &  0.758            & .SAS2*    & FN98 \\
NGC 7130   &  2.278            & .S..1P.   & GD98 \\
NGC 7582   &  1.111            & PSBS2..   & FN98 \\
Circinus   &  0.557            & .SAS3*.   & FN98 \\
MKN  477   &  1.779            & .S...*P   & HG97 \\
\hline
\end{tabular}

{\small 
\noindent CA96 --- Capetti et al. 1996 \\
\noindent CG97a --- Colina et al. , 1997a  \\
\noindent CG97b --- Colina et al. , 1997b  \\
\noindent FN98 --- Forbes et al. , 1998  \\
\noindent HG97 --- Heckman et al. , 1997  \\
\noindent GD97 --- Gonzalez Delgado, et al. , 1997a  \\
\noindent GD98 --- Gonzalez Delgado, et al. , 1998  \\
}

 From the above considerations, it seems reasonable to adopt 
 {$\rm L_{\rm FIR}/L_{\rm B}$} as an indicator of relative star formation 
 rate in Seyfert galaxies, and we will refer 
 $\rm L_{\rm FIR}/L_{\rm B} = 1/3$ as a threshold of enhanced 
 star formation.

\section{The Tully-Fisher Relation in Seyfert Galaxies}

Following the notations in Whittle (1992c), we replot the Tully-Fisher
relation, the total absolute blue magnitude $M_{\rm B} (total)$ 
against the normalized, inclination-corrected full rotation velocity 
$V_{\rm rot}^{\rm c}|Sb$ in Fig 1 for Whittle's Seyfert galaxies.
Fig. 1a is for $\rm L_{\rm FIR}/L_{\rm B} < 1/3$, and 
Fig. 1b for $\rm L_{\rm FIR}/L_{\rm B} \geq 1/3$.
The thick dot-dashed lines show the best fit to data, and
the thin dashed line illustrates the fit for normal Sb spirals from 
Rubin et al. (1985), given by,

\begin{equation}
M_{B} (total) = 5.84 - 10.23 * \log V_{\rm rot}^{\rm c}|{Sb}
\end{equation}

\noindent
where $V_{\rm rot}^{\rm c} = V_{\rm rot} /\sin i $, and $V_{\rm rot}^{c}| Sb 
= V_{\rm rot}^{\rm c} * 10^{0.053*(T-3)} $, and i, inclination angle, T is 
the numerical index of Hubble type. 

\noindent
$\rm L_{\rm FIR}/L_{\rm B} $ is computed with the relation

\begin{equation}
\log (\rm L_{\rm FIR}/L_{\rm B}) = -0.734 + 0.4 * B_{\rm T}^{\rm 0} - 0.4 * M_{\rm FIR}
\end{equation}

\noindent
 where $\rm M_{\rm FIR}$ and $\rm B_{\rm T}^{0}$ are apparent magnitudes at FIR and 
 blue band, and all the above basic data 
 are taken from the Lyon-Meudon Extragalactic Database 
 (LEDA) and RC3 (de Vaucouleurs et al. 1991). 

 The Kolmogorov-Smirnov tests indicate that Whittle's Seyferts with 
 $\rm L_{\rm FIR}/L_{\rm B} < 1/3$ obey the same Tully-Fisher relation as
 normal galaxies, shown in Fig 1a, while those with 
 $\rm L_{\rm FIR}/L_{\rm B} \geq 1/3$ ( in Fig 1b) show significant offset
 from the normal Tully-Fisher relation represented by thin dashed line.
 Considering our discussion on $\rm L_{\rm FIR}/L_{\rm B}$ in the previous
 section, the results shown in Fig 1 might imply that the host galaxies
 of Whittle's Seyferts with normal star-forming activities show the same
 properties as the normal spiral galaxies, while those with enhanced star
 formation do not. The reason is that the massive O, B stars, formed in the 
 enhanced episode of star formation, make a significant contribution to
 the galactic luminosities but not to masses (Rhee \& van Albada 1995).
 
 In order to examine the above arguments, we have constructed a new
 sample of Seyfert 2 galaxies from Veron-Cetty \& Veron (1996). The sample
 contains 52 sources, due to the limitation of basic data needed
 for the Tully-Fisher relation, which are extracted from LEDA and RC3.
 Fig 2a and 2b show the Tully-Fisher relation for this sample with
 the same definitions as those in Fig 1a and 1b.

 The first impression on Fig 2a and 2b, compared with Fig 1a and 1b,
 is that the new sample of Seyferts show much larger scatters. The
 main reason is that the Seyfert galaxies in Whittle sample shown
 in Fig 1 are constrained to those with better data quality (
 see Whittle 1992c). However, Seyfert 2 galaxies in the new sample
 are collected from all we can obtained without considering data
 quality. One example might be indicative.
 NGC 7319, one outlier in Fig 2a, has been in Whittle's large sample
 (Whittle 1992a), it was finally discarded in Whittle (1992c) in 
 analysing Tully-Fisher relation because of its large uncertainty
 in inclination, which is the main source of uncertainty in the 
 inclination-corrected
 full ratational velocity. We suspect that the other two outliers
 in Fig 2a, NGC 5347 and ESO 428-G14, have the same problem as
 NGC 7319. Statistically, the dot-dashed line in Fig 2a shows the best
 fit to the data, after discarding the three outliers. The 
 Kolmogorov-Smirnov test indicates that our new sample of
 Seyfert 2 galaxies with $\rm L_{\rm FIR}/L_{\rm B} < 1/3$ obey
 the same Tully-Fisher relation as the normal spiral galaxies.

 In constructing our new sample of Seyfert galaxies, we have not
 included any of Seyfert 1 sources, due to the unclear nature of FIR
 emission. Indeed, Maiolino et al. (1995) claimed that Seyfert 2
 galaxies experience higher level of star formation activities than
 Seyfert 1s. The analyses by Coziol et al (1998) for their PDS
 galaxies also derived a higher relative star formation rate for
 their PDS Seyfert 2 sample than for Seyfert 1s. 
 Gonzalez Delgado et al. (1997b) have listed  several Seyfert 1
 galaxies with detection of nuclear/circumnuclear star formation.
 Their relative star formation rates, $\rm L_{\rm FIR}/L_{\rm B}$,
 are 0.214, 2.790, 0.264, 0.104, and 0.257 for NGC 2639, NGC 7469, NGC 3227, 
 NGC 4639 and NGC 6814, respectively. 
 The majority of them show normal star formation activities,
 $\rm L_{\rm FIR}/L_{\rm B} < 1/3$. 
 It might be reasonable then to argue that the FIR emission in Seyfert
 1 galaxies  is different from that in Seyfert 2s. 

\section{Conclusions}

 We have discussed the effect of enhanced star formation in Seyfert 2 galaxies on
 the Tully-Fisher relation in this paper. The main results are (1) for nearly all 
 Seyfert 2 galaxies with detection of nuclear and/or circumnuclear star formation, 
 the ratio between FIR and blue luminosities is greater than 1/3, and (2) we confirm 
 that the offset of the Tully-Fisher relation in Seyfert 2 galaxies is  due
 to the enhanced star formation in the circumnuclear region.
 By use of $\rm L_{\rm FIR}/L_{\rm B}$ 
 as an indicator of relative star formation rate,
 we find that Seyfert 2 galaxies with normal star formation activities, 
 obey the same Tully-Fisher relation as the normal galaxies. 

\vspace*{5.mm}
\noindent
{\it acknowledgements.}  We have made use of data from the Lyon-Meudon 
 Extragalactic Database (LEDA) compiled by the LEDA team at the 
 CRAL-Observatoire de Lyon (France).
 This work is supported by a grant from the NSF of China, and a grant from 
 the Ascent Project of the State Science Commission of China.

\noindent
{\bf References}

\noindent Capetti A., Axon D.J., Macchetto F., Sparks W.B. \& Boksenberg A., 1996, ApJ, 466, 169.

\noindent Cid Fernander R. \& Terlevich R., 1995, MNRAS, 272, 423.

\noindent Colina L., Garcia-Vergas M.L., Mas-Hesse J.M., Alberdi A. \& Krabbe A., 1997a, ApJ, 484, \\\hspace*{5mm} L41.  

\noindent Colina L., Garcia Vargas M.L., Gonzalez Delgado R.M., Mas-Hesse J.M.,
          Perez E., Alberdi A. \& Krabbe A., 1997b, ApJ, 488, L71.

\noindent Combes, F., Prugeniel, P, Rampazzo, R., \& Sulentic, J.W. 1994, A\&A, 281, 725.

\noindent Condon J.J., Anderson M.L. \& Helou G., 1991, ApJ, 376, 95.

\noindent Coziol R., Torres C.A.O., Quast G.R. \& Contini T., 1998, astro-ph/9807136.

\noindent{de Vaucouleurs, G., de Vaucouleurs, A., Corwin Jr., H.G.,
        Buta, R.J., Paturel,\\\hspace*{5mm} G., \& Fougue, P. 1991,
        Third Reference Catalogue of Bright Galaxies, Springer
        \\\hspace*{5mm} Verlag, New York} (RC3). 

\noindent Devereux N.A. \& Young J.S., 1990, ApJ, 350, L25. 

\noindent Devereux N.A. \& Young J.S., 1991, ApJ, 371, 515. 

\noindent Devereux N.A. \& Scowen P.A., 1994, AJ, 108, 1244.

\noindent Devereux N.A., Price R., Wells L.A., \& Duric N., 1994, AJ, 108, 1667.

\noindent Devereux N.A., Jocoby G., \& Ciardullo R., 1995, AJ, 110, 1115.

\noindent Dultzin-Hacyan D., Moles M. \& Masegosa J., 1988, A\&A, 206, 95.

\noindent Feinstein C., 1997, ApJS, 112, 29.

\noindent Forbes D.A. \& Norris R.P., 1998, astro-ph/9804298.

\noindent Genzel R., Weitzel L., Tacconi-Garman L.E., Blietz M., Cameron M., 
    Krabbe A., Lutz D. \& Sternberg A., 1995, ApJ, 444, 129. 

\noindent Gonzalez Delgado R.M., Perez E. Tadhunter C., Vilchez J.M. \& 
          Rodriguez-Espinosa J.M. , \\\hspace*{5mm} 1997a, ApJS, 108,155.

\noindent Gonzalez Delgado R.M. \& Perez E., 1997b, ApJS, 108, 199.

\noindent Gonzalez Delgado R.M., Heckman T., Leitherer C., Meurer G., Krolik J., 
          Wilson A.S., Kinney \\\hspace*{5mm} A. \& Korattkar A., 1998, astro-ph/9806107.

\noindent Gu Q.S., Huang J.H., Su H.J., \& Shang Z.H., 1997, A\&A, 319, 92.

\noindent Heckman T., Blitz L., Wilson A.S., Armus L. \& Miley G.K., 1989, ApJ, 342, 735.

\noindent Heckman T., Krolik J., Meurer G., Calzetti D., Kinney A., etc, 1995, ApJ, 452, 549.

\noindent Heckman T.M.,Gonzalez Delgado R.M., Leitherer C., Meurer G.R., Krolik J., Kinney 
          A., Koratkar A. \& Wilson A.S., 1997, ApJ, 482, 114.

\noindent Helou G. \& Bicay M.D., 1993, ApJ, 415, 93.  

\noindent Huang J.H., Gu Q.S. Su H.J. et al., 1996, A \& A, 313, 13.

\noindent Kandalian R., 1997, IAU Sym., 184, in press.

\noindent Keel, W.C. 1993, AJ, 106, 1771

\noindent Kennicutt  Jr. R.C. \& Kent S. M., 1983, AJ, 88, 1094.

\noindent Kennicutt  Jr. R.C., 1998, ARA\&A, in press. (astro-ph/9807187)

\noindent Lonsdale Persson C.J. \& Helou G., 1987, ApJ, 314, 513.


\noindent Maiolino R., Ruiz M., Rieke G.H. \& Keller L.D., 1995, ApJ, 446, 561. 

\noindent Neff S., Fanelli M.N., Roberts L.J., O'Connell R.W., Bohlin R., 
          Roberts M.S., Smith A.M. \& Stecher T.P., 1994, ApJ, 430, 545.

\noindent Nelson C.H. \& Whittle M., 1995, ApJ, 465, 96.  

\noindent Norman C. \& Scoville N., 1988, ApJ, 332 124.

\noindent Oliva E., Origlia L., Kotilainan J.K., \& Moorwood A.F.M., 1995, A\&A, 301, 55.

\noindent Perry J. \& Williams R., 1994, MNRAS, 260, 437.

\noindent Rhee M.-H. \& van Albada T.S., 1995,
 IAP Astrophysics Meeting: Cosmic Velocity Fields, p89.

\noindent Rowan-Robinson  M., 1987, in Star Formation in Galaxies, ed. C.J. 
 Persson, U.S. Govt. Print. Off., Washington, DC, USA, p133.

\noindent Rodriguez-Espinosa J.M., Rudy R.J. \& Jones B., 1986, ApJ, 309, 76. 

\noindent Rodriguez-Espinosa J.M., Rudy R.J. \& Jones B., 1987, ApJ, 312, 555.

\noindent Rodriguez-Espinosa J.M. \& Perez Garcia A.M., 1998, ApJ in press

\noindent Romanishin W., 1990, AJ, 100, 373.

\noindent Ryder S.D. \& Dopita M.A., 1994, ApJ, 430, 142.

\noindent Terlevich R. \& Melnick J., 1985, MNRAS, 213, 841.

\noindent Terlevich R., Tenorio-Tagle G., Franco J. \& Melnick J., 1992, MNRAS, 255, 713.

\noindent Telesco C.M., Becklin E.E., Wynn-Williams C.G. \& Harper D.A., 1984, ApJ, 282, 427. 

\noindent Vaceli M. S., Viegas S. M., Gruenwald R \& Benavides-Soares P. 1993, PASP, 105, 875. 

\noindent Veron-Cetty M.P. \&  Veron P., 1996, A Catalogue of Quasars and Active Galactic
 Nuclei, 7th ed., ESO Scientific Report 17.

\noindent Whittle M., 1992a, ApJS, 79, 49.

\noindent Whittle M., 1992b, ApJ, 387, 109.

\noindent Whittle M., 1992c, ApJ, 387, 121.

\noindent Wilson A.S., 1988, A \& A, 206, 41.

\noindent Young J.S., Allen L., Kenny J.D.P., Lesser A. \& Rownd B., 1996, AJ, 112, 1903.

\clearpage

\noindent
{\bf \large Figure Captions:}

\noindent
Figure 1. The Tully-Fisher relation for Whittle's sample. (a) for Seyfert 
galaxies with $\rm L_{\rm FIR}/L_{\rm B} < 1/3$; and (b) for Seyfert galaxies
with $\rm L_{\rm FIR}/L_{\rm B} \geq 1/3$. The dot-dashed lines show the best
fit to data and dashed line is the fit to normal Sb spiral galaxies 
from Rubin et al. (1985).

\noindent
Figure 2. The Tully-Fisher relation for Seyfert 2 galaxies from Veron-Cetty
\& Veron (1996) sample. (a) for Seyfert galaxies with 
$\rm L_{\rm FIR}/L_{\rm B} < 1/3$; and (b) for Seyfert galaxies
with $\rm L_{\rm FIR}/L_{\rm B} \geq 1/3$. The dot-dashed lines show the best
fit to data and dashed line is the fit to normal Sb spiral galaxies 
from Rubin et al. (1985).

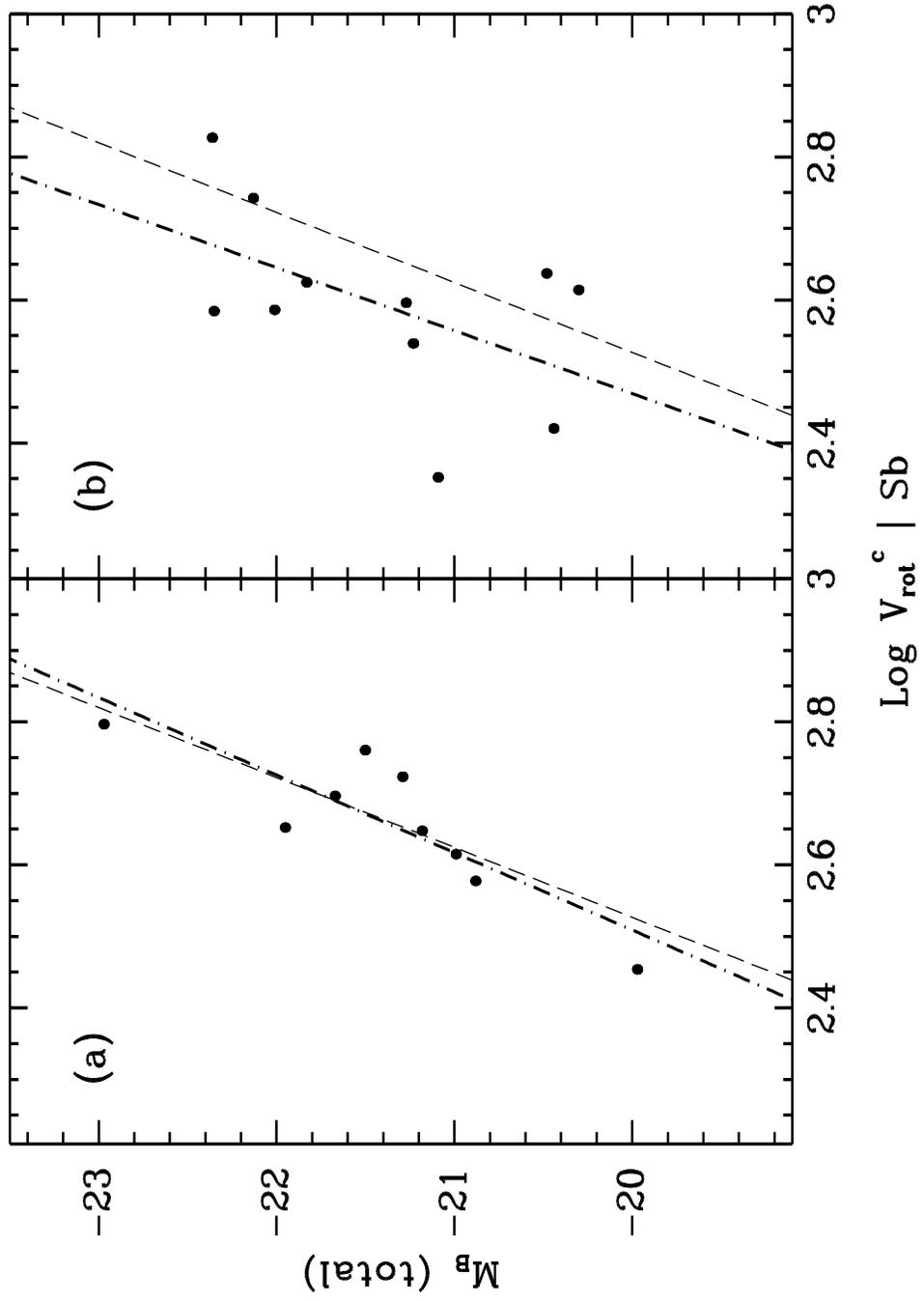
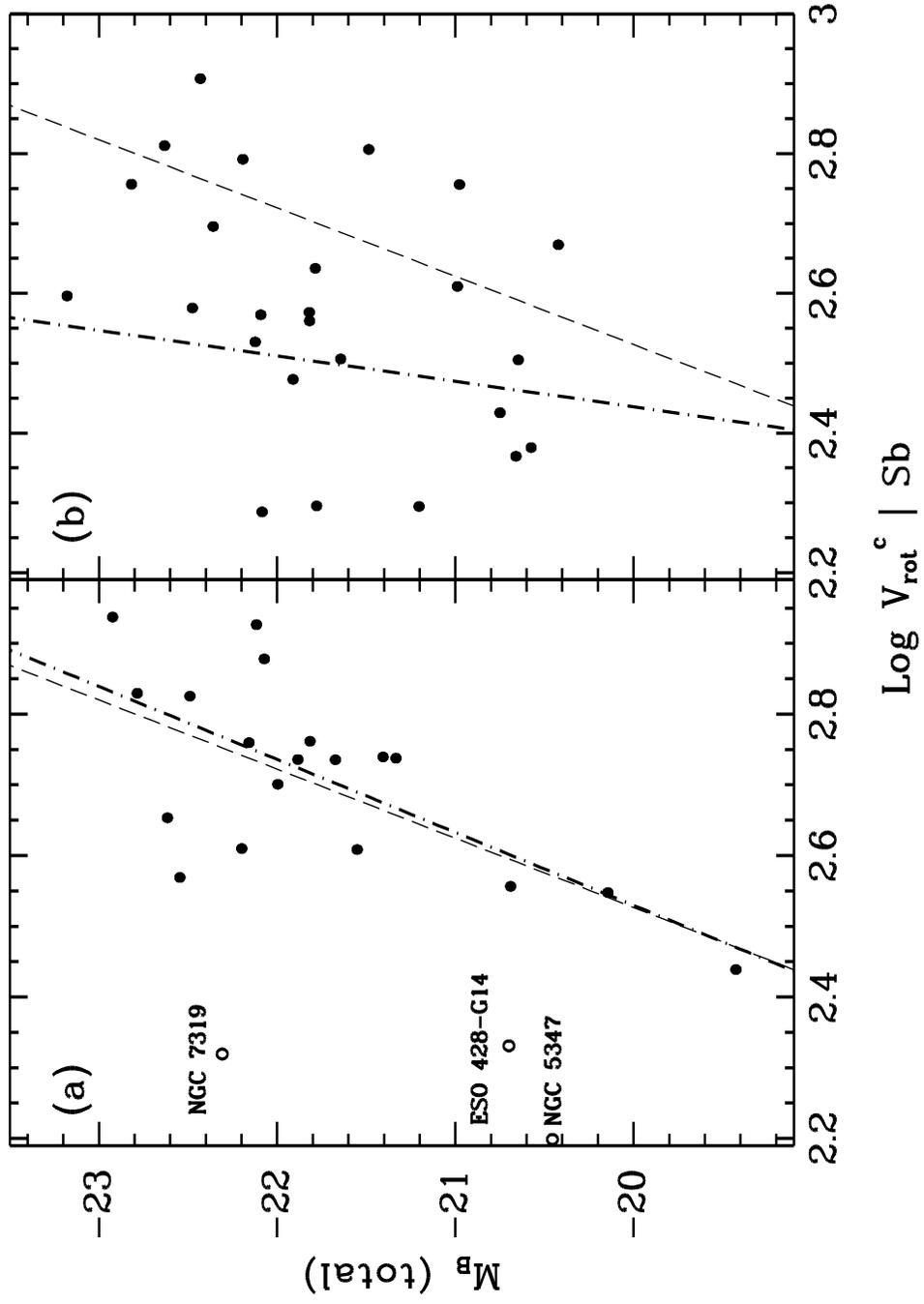
\end{document}